%% file: paper.tex
\documentstyle[preprint,aps,prb]{revtex}

\begin{document}
\draft
\preprint{LA-UR-97-1058}

\input zabst.tex

\input sect1.tex
\input sect2.tex
\input sect3.tex
\input sect4.tex
\input sect5.tex

\input zref.tex
\input zfigs.tex

\end{document}

%% file: zabst.tex
\title{\bf The onset of self-assembly}
\author{Shirish M. Chitanvis}
\address{
Theoretical Division, 
Los Alamos National Laboratory\\
Los Alamos, New Mexico\ \ 87545\\}

\date{\today}
\maketitle
\begin{abstract}
We have formulated a theory of self-assembly
based on the notion of local gauge invariance at the mesoscale.
Local gauge invariance at the mesoscale
generates the required long-range entropic forces responsible
for self-assembly in binary systems.
Our theory was applied
to study the onset of mesostructure
formation  above a critical temperature in estane, a diblock copolymer.
We used diagrammatic methods to transcend the Gaussian approximation
and obtain a
correlation length $\xi \sim (c-c^*)^{-\gamma}$, where
$c^*$ is the minimum concentration below which self-assembly is impossible, 
$c$ is the current concentration, and
$\gamma$ was found numerically to be fairly close to $2/3$.
The renormalized diffusion constant vanishes as the critical
concentration is approached, indicating the occurrence of
critical slowing down, while the correlation function remains finite
at the transition point.
\end{abstract}

%% file: sect1.tex
\section{Introduction}

Microphase separation is the tendency in certain mixtures 
such as amphiphilic fluids and diblock copolymers,
of one of the components to form mesoscale aggregates of the size of 
$\sim$ 100 $\AA$.
Such self-assembly is to be distinguished from the phenomenon of
nucleation in single-component fluids.
Nucleation is the precursor of a phase transition, and as such
indicates an instability.
Self-assembly on the other hand indicates the ability of a given
mixture to grow {\it islands} of one of the components to the
size of a couple of hundred Angstroms, and then stabilize the growth,
so that phase separation occurs only on a mesoscale, rather than
a macroscale.
Above a certain temperature $T^*$, mesoscale structures are
formed in a random fashion, as long as the concentration of
the self-aggregating component is greater than some minimum value $c^*$.
As the concentration is increased continuously above $c^*$,
the self-assembling systems first
form spherulitic structures, changing to fibrillar and then lamellar
structures.\cite{dobrynin}
This is inferred experimentally using small-angle (X-ray or
neutron) scattering.
The correlation functions in k-space obtained from such experiments
display a peak around some wave-vector indicating the average
spacing between these islands.\cite{bonart,chandler}
The width of the peak represents the spread in the average
spacing of these islands.
Below $T^*$, the mesoscale aggregates form regular arrangements
(e.g. hcp, fcc, etc) via a first order transition.\cite{bates,chow}
The regularity of these lattices can be inferred from small-angle scattering
experiments which display harmonics of the main peak.\cite{bates}

It is generally believed that self-assembly in mixtures 
is due to the competition between
the tendency of the components to phase separate on a macroscale,
and a long-range entropic (statistical) force  caused by the presence
of chemical bonds linking the components in the mixture.\cite{dobrynin}
In the case of amphiphilic mixtures, it is the surfactant molecules
which provide the glue which allows mesoscale segregation to occur.
In the case of diblock copolymers, end-groups on the two species
create inter-specie bonds, thereby playing the role of a surfactant.
A molecular-level description of mesoscale structures (micelles) in liquids,
and aggregates in copolymers that are a couple of hundred angstroms in 
size is a challenging problem.  At this scale, raw simulations which 
begin at the molecular level are simply impossible to perform
for realistic molecules with the
current computational technology.

Parallel field theoretic efforts in both amphiphilic fluids as well as 
diblock copolymers have been developed over the years
to provide an understanding
of microphase separation.
\cite{chandler,leibler,degennes,ohta,stillinger,woo}
We will show in this paper that the principle of local gauge invariance 
with respect to the $SO(2)$ group
can be applied successfully to unify the above theories with a 
common thread, and furthermore, to derive a generalization of these mesoscopic
theories of self-assembly.\cite{shirish}
We have interpreted the
gauge fields we obtain as giving rise to statistical correlations between
concentration fluctuations.
These statistical correlations could be thought of as effective interactions
which arise at the mesoscale from the underlying Coulombic interactions
at the molecular level,
between the components of the mixture.
While the use of local gauge invariance\cite{YM} is quite
well established in particle physics, its usefulness in
settings other than quantum field theory (QFT) is 
appreciated only under rare circumstances.\cite{edelen}
We note that while the {\it dynamical} use of local gauge invariance is
novel at the mesoscale, gauge theory has been used routinely in the past
to classify defects in condensed matter physics.\cite{mermin}
Our theory is applicable to diblock copolymers, oil-water-surfactant
mixtures, and in general any self-assembling system, e.g.
binary alloys.\cite{ducastelle}

A further importance of our paper lies in the fact that we 
have gone beyond
the Gaussian approximation or the Mean Field Approximation (MFA) 
used conventionally in meso-scale theoretical investigations.
\cite{leibler,ohta}
While the MFA may be a reasonable approximation to study
self-assembling systems far from phase transitions, it
is obvious that one must necessarily go beyond the MFA
or the Gaussian approximation in order to properly study
the onset of self-assembly.
To be more precise, we point out that investigations of
the onset of self-assembly (as the composition is varied)
in the literature\cite{leibler,stillinger,woo} yield a correlation length
diverging with the square-root signature of the MFA.
Experimental observations cited by Woo et al\cite{woo} suggest that the
true exponent is larger than ${1\over2}$.
We will take seriously in this paper the quantitative suggestion of Woo et al
that there is a need to go beyond the MFA or the Gaussian approximation
to study the onset of self-assembly.

There have indeed been investigations
in the past where Renormalization Group (RG) techniques have been used
to study the first order transition from a disordered to an ordered phase
in copolymers, as the temperature
is varied.\cite{bates,decruz,stuhn,barrat,dobrynin}
But the onset of the self-assembly of mesoscopic structures 
into a random arrangement above a critical temperature, as the
concentration of one of the components of the binary mixture is
varied, is an issue that has not been addressed theoretically in much detail
beyond the Gaussian approximation.
Our investigation reveals that this transition is analogous to the
critical point in phase transition theory, in that
the correlation length diverges as a $2/3$ power law,
and the diffusion constant goes to zero, implying critical slowing down.
But the correlation function itself does not diverge.
In this sense, we are investigating a Lifshitz point.\cite{barbosa}
Our detailed calculations apply specifically to estane, a diblock copolymer.
However, one may invoke universality arguments to argue that our
results are applicable more generally.

Our theory
represents a generalization of the $\phi^4$ field theory proposed
by Landau and Ginzburg to study phase transitions.
While our theory is slightly similar to the
standard Landau-Ginzburg theory
in that they are both nonlinear
and deal with an order parameter, it is clear that there are some
major differences.
First of all, our theory is nonlocal in character.
Secondly, the nonlinear term in our theory not only
contains a cubic term (in addition to a quartic term),
which arises naturally from 
an expansion around the average value of the fields, but the
nonlinear term also contains derivatives of the concentration.
The derivative form of the nonlinear coupling is dictated by the
fact that ours is a gauge theory, in which covariant derivatives
are defined.
We have used this theory to investigate the onset of
self-assembly in estane, a diblock copolymer,
and we found that the correlation length diverges with a power
which is fairly close to the universal value of (2/3).
We also found that the renormalized diffusion constant goes
to zero as the minimum concentration $c^*$
(below which self-assembly is imposible) is approached.

We foresee a rich variety of applications of our approach to other
questions regarding diblock copolymers, such as 
their viscoelastic properties.
We also foresee investigations of time-dependent phenomena in copolymers,
such as detailed studies of critical slowing down at the onset of self-assembly.

%% file: sect2.tex
\section{The gauge theory}

The starting point of our mesoscale theory is
an internal energy functional which is quadratic in the gradient of
a two-dimensional vector.
For the moment, we will
consider isolated systems, so that the 
quantity that is conserved is the internal energy.\cite{callen}
We will shortly consider entropy effects as well.
Consider the following form for the energy functional:

\begin{equation}
\beta U_o =  \beta \int u_0 ({\bf c}({\bf s})) {{\rm d}}^3 s 
\label{2.1}
\end{equation}
\begin{equation}
\beta = {{1}\over {k T}}
\label{beta}
\end{equation}

\begin{equation}
\beta u_o ({\bf c}({\bf s}))  =  \left({g\over2}\right) 
      {{\partial {\bf c}^t({\bf s})}\over{\partial s_i}}
                          {{\partial {\bf c}({\bf s})}\over{\partial s_i}}
\label{2.2}
\end{equation}

where $t$ indicates a transpose, 
repeated indices are summed over, and,

\begin{equation}
{\bf c}({\bf s}) \equiv \left( \matrix{ c_h ({\bf s}) \cr c_s ({\bf s})} \right)
\label{2.3}
\end{equation}

In the above equations, ${\bf s}$ is a dimensionless co-ordinate variable,
$k$ is Boltzmann's constant, $T$ is the temperature, $c_h$ is the number
concentration of the first type of specie, and 
$c_s$ is the number concentration
of the other specie in a binary mixture.  
The concentrations are normalized 
to the total number concentration.
The constant $g$ is essentially a dimensionless diffusion constant.
Such energy functionals have been considered over many years as
contributing to the total internal energy of binary mixtures.
\cite{stillinger,cahn}
We will use this from as our starting point to generate a more
complete energy functional using gauge invariance.

From Eqns.(1)-(4) we see that $u_o$ is invariant under
global rotations of the vector ${\bf c}$.  These are rotations in
two dimensions, and the appropriate group to consider is $SO(2)$.
The physical origin of this group can be traced back to the fact that the
quadratic (positive, semi-definite) form of the energy density (Eqn.(3))
is dictated by expanding the internal energy around a minimum, in a 
Landau-like fashion.
The form of the energy density contains gradient operators,
which permits us to perform $SO(2)$ transformations around not
just the origin in $(c_h,c_s)$ space, but around any arbitrary
fixed vector in this space.
In particular, we shall use $SO(2)$ around the vector defined
by the average concentration of each specie viz., $(c_h^o,c_s^o)$.
This is a natural representation for our system, since our final goal
is to study self-assembly in binary systems, characterized by
local, mesoscale fluctuations around the average concentrations.
$SO(2)$ transformations of these fluctuations demands that 
${c'}_h^2 + {c'}_s^2 $ = constant,
where ${c'}_h$ and ${c'}_s$ denote deviations of the specie concentrations
around their averages.
Thus $SO(2)$ transformations can cause the components of 
$({c'}_h, {c'}_s)$ to become negative.
But this is acceptable, since concentration fluctuations
around the average can indeed be negative or postive,
as long as the total concentration for each specie does
not become negative (see Eqn.(11)).
In what follows we shall be tacitly performing local $SO(2)$
transformations around the average concentration vector $(c_h^o,c_s^o)$,
culminating in Eqn.(13), which is a central result in our paper.

Our physical motivation for seeking local gauge invariance 
of ${c'}_h^2 + {c'}_s^2 $ under
$SO(2)$ is the same as that of Yang and Mills\cite{YM}, and 
in quantum electrodynamics, where 
one observes the invariance of the noninteracting Lagrangian,
which is
bilinear combinations of the fields,
under certain global transformations.
One then demands covariance of the theory when these symmetry
operations are {\it local} i.e., when the transformations are
space-time dependent.
A reason for this, as given by Yang and Mills, is that one can 
now freely interchange between the fields as one moves through
space and time, while leaving the physics covariant.
It is intuitively clear that such is the case in our problem,
where chemical connections between the two species n our system
allows for an admixture of the two components, rather than
permitting a complete phase separation to occur on a macroscale.
Thus, instituting local gauge invariance under $SO(2)$ in our
binary mixture is equivalent to allowing interactions between the components.
Beyond this initial motivation, it is equally important to show
that the result of local gauge transformations of $u_o$ lead to
physically significant results as epitomized by Eqn.(13).

We remark in passing that $u_o$ is also invariant under the translation
group $T_2$, where we consider the transformations 
${\bf c} \to {\bf c} + \bf a$.
Based on the work of Edelen\cite{edelen} in solid mechanics,
we believe that seeking local gauge invariance of $u_o$ under $T(2)$
may lead to a study of defects in our system.

Following Yang and Mills,\cite{YM} local gauge invariance of $u_o$
under $SO(2)$ motivates us to define new fields $\bf b$, which have 
invariance properties appropriate to $SO(2)$.  
We define a covariant derivative 
${\partial\over{\partial s_i}} \to ({\partial\over{\partial s_i}} +
                                         q \tau b_i)$,
where $\tau$ is the generator of $SO(2)$, $q$ is a `charge', or
equivalently, a coupling constant, and the $b$-fields are analogs
of the magnetic vector potential in electrodynamics. 
These $b$-fields give rise
to effective interactions between the hard and soft segments of
estane.
These effective interactions are to be thought of as arising form
the underlying electrostatic interactions between molecules, monomers,
etc.
The energy functional
for the $b$-fields is defined 
$\grave {\rm a}$ la Yang and Mills, via the minimal
prescription.  
With this, our original internal energy density is transformed
into:

\begin{equation}
\beta u_o \to \beta u = \beta u_o + \beta u_{int} + \beta u_{YM} 
\label{}
\end{equation}

where $u_{int}$ refers to the interaction energy density, and $u_{YM}$
is the energy density associated with the Yang-Mills $b$-fields alone.
Equivalently, we may define the total energy functionals associated
with these energy densities:

$$
\beta U_o \to \beta U = \beta U_o + \beta U_{int} + \beta U_{YM},
$$

where

\begin{equation}
\beta u_{int} = J_i ({\bf c}) b_i ({\bf s}) +
           b_i ({\bf s}) f({\bf c}) b_i ({\bf s}) 
\label{}
\end{equation}

with

\begin{equation}
J_i ({\bf c}) = \left({1\over2}\right) 
      q g \left( {{\partial {\bf c}^t({\bf s})}\over{\partial s_i}} 
                \tau {\bf c} ({\bf s}) +
                  {\bf c}^t ({\bf s}) \tau^t
                          {{\partial {\bf c}({\bf s})}\over{\partial s_i}}
                 \right)
\label{}
\end{equation}

\begin{equation}
f ({\bf c}) = \left({1\over2}\right) g q^2 {\bf c}^t ({\bf s}) 
                                            {\bf c}({\bf s});
\label{}
\end{equation}

$\tau$ is 
given by:\cite{sudarshan}

\begin{equation}
\tau = \pmatrix{0&-1\cr1&0}
\label{}
\end{equation}

From the above equation, it can be shown that
$$
J_i ({\bf c}) =
q g \left( {{\partial {\bf c}^t({\bf s})}\over{\partial s_i}}
                \tau {\bf c} ({\bf s})
\right)
\eqno(9a)
$$

We need one more definition for completeness:

\begin{equation}
\beta u_{YM} = \left({1\over{4}}\right)~
\left( {\partial b_i\over \partial s_j}-{\partial b_j\over \partial s_i}\right)
\left( {\partial b_i\over \partial s_j}-{\partial b_j\over \partial s_i}\right)
\label{}
\end{equation}

This equation can be cast into the following form:

$$
\beta u_{YM} = -\left({1\over{2}}\right)~b_i {\nabla}^2 b_i
\eqno(10a)
$$

Eqn.(10a) is obtained via an integration by parts, in the transverse gauge.
Since we are dealing with an Abelian gauge theory, it is permissible to
insert this transverse gauge manually, without resorting to the formal
machinery of Faddeev and Popov.

Note that we are utilizing a non-relativistic version of the
Yang-Mills procedure, since we are only concerned with time-independent
problems.
Furthermore, since we are concerned with rotations in two-dimensional
space, there is only a single generator for the group $SO(2)$ (see Eqn.(9)),
so that the resulting functional is only quadratic and not quartic
in the {\it b}-fields.

It is important to emphasize that the usual application of the Yang-Mills
procedure in QFT implies the existence of fundamental
interactions.  In our case, we are applying the principle of local
gauge invariance at the $mesoscale$.  Consequently, we do not expect to
discover any new fundamental interactions by using gauge invariance.
Rather, we interpret the new $b$-fields as yielding correlations between the
concentration fields.  As proof of this, we will show shortly that
our approach leads to a generalization of the theories of Stillinger
and Leibler, where correlations were invoked on physical grounds to
describe mesoscale structures. 
These correlations could also be thought of as effective interactions,
which arise at the mesoscale from the underlying electrostatic
interactions between molecules.
We will not address the question of how one can make a connection with
molecular scale properties in this paper.

Our approach is analogous to the Landau-Ginzburg theory of 
superconductors in magnetic fields.\cite{landau}
In the Landau-Ginzburg theory, the energy functional involving a
complex order parameter is gauged with
respect to the $U(1)$ group.  
This permits a successful treatment of
a superconductor in a magnetic field, and even permits a classification of
superconductors.
We have gone further in our theory, and invoked gauge invariance to
study correlations that develop at the mesoscale.  In our theory,
there is no external magnetic field to consider.

The partition function we need to evaluate is now:

\begin{equation}
Q = \int \prod_{\alpha=h,s}{{\cal D}c_\alpha} 
            ~\theta\left(c_\alpha\right) 
                      \prod_{k=1,3}{{\cal D}b_k}
                               ~exp-\beta (U_o+U_{int}+U_{YM})
\label{}
\end{equation}

Equation (11) is a functional integral,
where the step functions denoted by $\theta$ imply that we must 
restrict integration to positive semi-definite values of
the fields.

Since the $b$-fields appear only quadratically in the above
functional, it is straightforward to integrate over them, and obtain
an effective internal energy functional involving only ${\bf c}$.
The result is:\cite{kaku}

\begin{equation}
\beta U_{eff} = \beta U_o + \beta \Delta U_{eff} =
        \beta U_o -{1\over{4}} \int {\rm d}^3s \int {\rm d}^3 s'~
                        J_i({\bf c}({\bf s}))~
         \left(
            {1\over{f({\bf c}({\bf s})) -
                              {1\over{2}}\nabla^2
                   }
            }\right)_{{\bf s},{\bf s}'}~
                                   J_i({\bf c}({\bf s}'))
\label{}
\end{equation}

Note that in doing so, we have ignored an overall trivial normalization 
constant that appears in the evaluation of the partition function $Q$.
This is permissible, as this factor cancels during the evaluation of 
averages of observable quantities.

To see the connection between this rather complicated functional and 
the older theories, we expand the second term on the right hand side of 
Eqn.(12) around the average concentrations of the two species
($c_h^0$ and $c_s^0$) that
appear in our theory, and retain only quadratic terms.
The result is:

\begin{equation}
\beta U_{eff} \approx \beta U_o
  -\left({\Omega\over 2}\right) \int {\rm d}^3 s~{\bf {c'}}^t({\bf s}) {\cal S}
                                                         {\bf {c'}}({\bf s})
     + \left({\Gamma \over {2 \pi}}\right) \int {\rm d}^3 s \int {\rm d}^3 s'~
                     {\bf {c'}}^t({\bf s}) {\cal S} {exp(-\sqrt{2 \tilde g}
                                        \vert {\bf s} - {\bf s}' \vert)\over
                                              {\vert {\bf s} - {\bf s}' \vert}
                                           }
                                                 ~   {\bf {c'}}({\bf s}')
\label{gauss}
\end{equation}

where $\Omega = g^2 q^2 $, $\tilde g = 
          \left({g\over2}\right) q^2 ((c_h^0)^2+(c_s^0)^2)$,
$\Gamma=q^2 g^2 \tilde g $,
and the primes on $\bf c$ denotes deviations of the specie concentrations
from their averages.
The matrix $\cal S$ is defined thusly:

\begin{equation}
\cal S = \pmatrix{(c_s^0)^2&-c_h^0 c_s^0\cr-c_h^0 c_s^0&(c_h^0)^2}
\label{}
\end{equation}

First of all, we see that by expanding around the average value of the 
fields, we are in essence considering the effect of local $SO(2)$
invariance on the correlations which develop between the
fluctuations of the fields.
Secondly, we notice that if we set $\tilde g \to 0$, we recover
a model very similar to that of Stillinger and Leibler.
$\Omega$ is the Flory-Huggins parameter, and 
represents the immiscibility of the two components in our mixture.
It is prescribed automatically via gauge theory, as long as $g$ is
known.  
The non-local term in Eqn.(13) gives rise to correlations which tend
to counteract the effect of $\Omega$.
This frustration is responsible for the formation of
mesostructures.
From the definition of $\cal S$, we see that we have retained in our model
the notion of {\it (pseudo) electroneutrality} emphasized by 
Stillinger\cite{stillinger}
and Chandler {\it et al}.\cite{chandler}
Finally, we note that in general $\tilde g$ is not zero, so that
we have a screened Coulombic correlation appearing in the second term of 
Eqn.(13).
In this sense, Eqn.(13) may be viewed as being similar to the
random phase approximation (RPA) applied to the full functional
given by Eqn.(12).
Note that for small deviations ($c_h',c_s'$) from the corresponding
concentration averages, the step functions of Eqn.(11) have a 
negligible effect.

Equation (13) is one of the main results of our paper.
It shows that Leibler and Stillinger's theories may be understood
in the context of gauge theories.
Equation (13) gives credence to the notion that gauge theoretic
ideas may be valid at the mesoscopic level.

Equation (13) was obtained by expanding fields around their
average values.  In this sense we have broken the symmetry of our
system.
Combined with gauge invariance, then, we get a Yukawa-type screened
potential.
This effect may be interpreted by saying that the gauge fields
have acquired a mass.
In this sense, Eqn.(13) exhibits the Higgs phenomenon.\cite{higgs}

Before we can compare our theory with experimental data, we
need to consider the fact that our system is not really
isolated, and may be in contact with an energy reservoir, 
perhaps as it is being acted on by mechanical forces
in a stress experiment.
For a system in contact with an energy reservoir, the 
quantity that is conserved is the Helmholtz free energy\cite{callen}
$A = U - ST$, where $S$ is the entropy of the system.
The entropy of our system will be written in the usual form:

\begin{equation}
-{S\over k} = \int {\rm d}^3s~ (c_h({\bf s}) {\ln}(c_h({\bf s})) +
                                c_s({\bf s}) {\ln}(c_s({\bf s}))
                             )
\label{}
\end{equation}

with this, our theory is formally complete.

While our theory has been able to reproduce the older theories of
self-assembly of Leibler, Stillinger and Chandler, we believe that
the importance of our approach lies in the fact tht it provides a
natural way to go beyond the Gaussian approximation and the MFA
used conventionally in meso-scale investigations.
By this we mean that our effective functional can be expanded
in an infinite series beyond the Gaussian approximation.
While the MFA is a reasonable approximation to study self-assembling
systems far from phase transitions, it appears obvious that one
must necessarily go beyond the Gaussian approximation in order to
properly study phase transitions, e,g., the onset of self-assembly.

%% file: sect3.tex
\section{Going Beyond the Gaussian Approximation}

To see what lies beyond the Gaussian approximation, it is
convenient to 
invoke incompressibility,
so that we can cast 
the Helmholtz free energy solely in terms of the concentration of
$c_h$  the concentration of one of the species in our binary system 
(with the average concentration of that specie subtracted from it).
It is important to point out that the condition of incompressibility is
to be imposed after the starting functional $U_0$ has been gauged.
The condition of incompressibility is to be accounted for during
the evaluation of the partition function for the system.

\begin{equation}
A = \int {\rm d}^3s~a({\bf s}) \equiv U_o + \Delta U_{eff} - S T
\label{}
\end{equation}
\begin{eqnarray}
&&\beta a({\bf s}) = \beta a_o({\bf s}) + \beta \Delta a({\bf s}) \\
&&\beta A = \beta A_o + \beta \Delta A \\
&&U_o = \int {\rm d}^3s~g ({\bf \nabla}c_h({\bf s}))^2 \\
&&\beta \Delta U_{eff}  =
    -\left({{g^3 q^4}\over8}\right)
       \int {\rm d}^3s' \int {\rm d}^3s \vec \nabla_{s'} c_h({\bf s'}) \cdot
         \left(\hat{g_o}
          \left(\gamma c_h + c_h^2\right)
            \hat{g_o} \right)_{{\bf s'},{\bf s}}
                         \vec \nabla_{s} c_h({\bf s}) \\
&&-{S\over k} = \int {\rm d}^3s~ (c_h({\bf s}) {\ln}(c_h({\bf s})) +
                                (1-c_h({\bf s}) {\ln}(1-c_h({\bf s}))
                             )
\label{}
\end{eqnarray}
where $\beta = 1/k_B T$, and

\begin{eqnarray}
\beta a_o({\bf s}) &&\approx 
\left(1/(c_h^0 (1-c_h^0))\right) c_h({\bf s})^2
        + g ({\bf \nabla}c_h({\bf s}))^2 \nonumber\\
           && -\left({\Omega\over 2}\right) c_h^2({\bf s}) +
               \left({\Gamma\over{2 \pi}}\right) c_h({\bf s})
                                 \int {\rm d}^3s'~
                            {exp(-\sqrt{2 \tilde g}
                          \vert {\bf s} - {\bf s}' \vert)\over
                        {\vert {\bf s} - {\bf s}' \vert}
                   }~  c_h({\bf s}') 
\label{m1}
\end{eqnarray}

\begin{eqnarray}
&&\beta \Delta A \approx 
   \alpha \int^* {\rm d}^3s \vec \nabla_s c_h({\bf s}) \cdot
          \left(\gamma c_h({\bf s}) + c_h({\bf s})^2\right)
                         \vec \nabla_{s} c_h({\bf s})
\label{m2}
\end{eqnarray}

where $\Omega = g^2 q^2 $, $\tilde g = 
          \left({g\over2}\right) q^2 ((c_h^0)^2+(c_s^0)^2)$,
and 
$\Gamma=q^2 g^2 \tilde g $, $g$ is essentially a dimensionless
diffusion constant, and $q$ is a pseudo-charge that arises out
of our gauge theoretic considerations,
$c_h^0$ and $c_s^0$ are the average concentrations of the two
individual species in our system, 
and

\begin{eqnarray}
&&\hat{g_o} = \left(f(c_h({\bf s})) - {1\over 2} \nabla^2 \right)^{-1} \\
&&f(c_h({\bf s})) = {1\over 2} g q^2 (c_h^2({\bf s})+(1-c_h({\bf s}))^2) \\
&&\alpha = {g\over4 ((c_h^0)^2+(c_s^0)^2)} \\
&&\gamma = 4 \left(c_h^0 -1/2 \right) 
\label{cc}
\end{eqnarray}

The competition between the Flory-Huggins 
separation parameter $\Omega$
and the attractive non-local term 
in $a_o$ (Eqn.(\ref{m1})) gives rise to the formation of
mesostructures.
In obtaining Eqn.(\ref{m2}), we have ignored terms linear in the fields, and
constant terms, as they do not contribute to the density-density
correlation function.
In Eqn.\ref{m2}, we have looked for higher order corrections to
$\Delta U_{eff}$.
We have ignored cubic and quartic contributions which come from
the entropy term, as our
diagrammatic estimates indicate that they are negligible.

The form of $\Delta A$ in Equation \ref{m2} is a local
form. 
The local form is obtained by 
retaining only the lowest order non-linear terms
in an expansion of the full nonlocal form of the interaction term.
It is a reflection of
the fact that the full form of the nonlocal interaction 
term is screened on a length scale ~$1/\sqrt{2 \tilde g}$.
The asterisk on the integral in Equation \ref{m2} indicates that a
cutoff in momentum space is to be used in the short wavelength limit,
$k_{max} = \sqrt{2 \tilde g}$.
There would be no need for a cutoff if the full form of $\Delta A$
were to be used.
We note that Equations \ref{m1} and \ref{m2}
represent a generalization of the $\phi^4$ field theory proposed
by Landau and Ginzburg to study phase transitions.

We will now apply this gauge theory to the onset of self-assembly in
estane, a diblock copolymer.
Estane is composed of hard segments of polyurethane, and soft strands
of polyester.
The hard segments display microphase separation on the scale of $\cal O$(100)
Angstroms.
It is appropriate to consider an approximation
to the form of $a_o$, which ensures that Porod's law is satisfied
in the small wavelength limit.\cite{woo}

\begin{equation}
\beta \hat a_o({\bf k}) \approx \hat c^*_h({\bf k}) 
                \left( a + g' k^2 + \Gamma' k^4 \right)
                        \hat c_h({\bf k})
\label{gau}
\end{equation}

where:
$g'= g/(1-1/2((c_h^0)^2 + (c_s^0)^2)$, 
$\Gamma' = 1/(2 q^2 ((c_h^0)^2 + (c_s^0)^2))$,
and 
$a = 1/(2 c_h^0 (1-c_h^0))$.

The term in parantheses in
equation \ref{gau} represents the inverse of the structure factor
(Fourier transform of the density-density correlation function) in the
Gaussian approximation.
Following the formulation of our gauge theory, $g' > 0$.
As such, the Gaussian approximation has to be improved upon before
seeking agreement with experiment.
Unless $g'$ gets remormalized to a negative value, the structure
function will not yield a peak at some non-zero value of the
wave number, which would characterize a micro-phase separated system.
In what follows, we shall drop the subscript $h$ which appears on
the field $c$.
The partition function is defined as
($J$ is an auxiliary field):

\begin{equation}
Q\left[J\right] = 
      \int {\cal D} c \exp-\beta \left(a_o({\bf s}) + \Delta a({\bf s}) 
                                     + 2 J({\bf s}) c({\bf s}) 
                                \right)
\label{Q}
\end{equation}

We can now use standard perturbation techniques\cite{ramond} to develop
a series expansion for corrections to the Gaussian approximation.
As such, it is meaningful to make certain that the dimensionless
coupling constant $\alpha$ defined in Equation \ref{cc}
is less than one.
On the other hand, it
is well-known\cite{binney} that such series are asymptotic in nature.
Thus $\alpha < 1$ is not a panacea.
We define the two-point correlation function as follows:

\begin{equation}
S({\bf x_1},{\bf x_2}) = \left({1\over2}\right) 
              \left[
                  {\delta\over\delta J({\bf x_1})}
                                         {\delta\over\delta J({\bf x_2})} 
                                               \ln Q\left[J\right]
              \right]_{J=0}
\label{S}
\end{equation}

Figure 1 gives a pictorial representation of the two terms in the definition of
$\Delta a$.
It is clear that each of these two interaction terms yields a separate
perturbation series.
In addition, there will be a series formed out of the cross-terms as well.
There are no cross-terms
up to the 2-loop level.
The cubic interaction term first
yields non-vanishing contributions 
in second order
perturbation theory (Figures 2a-2b).
The quartic term yields non-zero contributions at the one-loop level
(Figures 3a-3b).
We have verified explicitly that {\it all} other (asymmetric) diagrams arising
up to the 2-loop level add up to yield a null contribution.
Similar cancellations are also obtained in theories of dendritic growth.
\cite{janssen}
In our calculation, the two series arising out of each
of the two interaction terms were evaluated only to the first 
non-vanishing order.

Figure 2a (tadpole) renormalizes $g'$:
\begin{equation}
\delta g'(1) = -\left({2\over {a \pi^2}}\right) \alpha^2 \gamma^2 
         \int_{0}^{k_{max}} {\rm dk} 
               {k^4\over \left( a + g' k^2 + \Gamma' k^4 \right)}
\label{tadpole}
\end{equation}

where $k_{max}=\sqrt{2 \tilde g}$.

This tadpole diagram is crucial in helping us achieve agreement with
experiment.
It dominates the contributions from Eqns.\ref{ss} and\ref{scoop}.
Note that it has a sign opposite that given by Eqn.\ref{scoop} below.
Without Eqn.\ref{tadpole}, the renormalized structure function
would not yield the characteristic peak in scattering data. 
This tadpole diagram is reminiscent of the standard (Hartree)
tadpole diagram in many-body physics.
The physical importance of this diagram is as follows.
The bare diffusion constant $g$, if left unregulated, would tend
to smooth out concentration gradients in our system.
It is the role of the screened "Coulomb" interactions, having
a statistical origin, which are responsible for self-assembly.
And it is up to these interactions to counteract the smoothing
tendency of the diffusion term.
This is accomplished as described above, by the tadpole diagram,
which renormalizes the bare diffusion constant so that the 
renormalized diffusion constant is less than or equal to zero.

Figure 2b yields two terms in leading order, 
one which renormalizes $g'$, and the 
other which renormalizes $\Gamma'$:

\begin{eqnarray}
&&\delta g'(2) = -\left({4\over {3 \pi^2}}\right) \alpha^2 \gamma^2 
                        \int_{0}^{k_{max}} {\rm dk} 
               {k^4\over \left( a + g' k^2 + \Gamma' k^4 \right)^2}
                                        \nonumber\\
&&\delta \Gamma'(1) =
  -\left({2\over \pi^2}\right) \alpha^2 \gamma^2 \int_{0}^{k_{max}} {\rm dk} 
           {k^2\over \left( a + g' k^2 + \Gamma' k^4 \right)^2}
\label{ss}
\end{eqnarray}

The renormalization of $g$ caused by this diagram is dominated
by the contribution of the tadpole diagram discussed above.

Figure 3a (1-loop) renormalizes $g'$:

\begin{equation}
\delta g'(3) = \left({2\over \pi^2}\right) \alpha \int_{0}^{k_{max}} {\rm dk} 
                 {k^2\over \left( a + g' k^2 + \Gamma' k^4 \right)}
\label{scoop}
\end{equation}

Figure 3b (1-loop) renormalizes the constant $a$ in Equation \ref{gau}.
A closed form expression for the contribution from this diagram is:

\begin{equation}
\delta a(1) = \left({2\over \pi^2}\right) \alpha \int_{0}^{k_{max}} {\rm dk} 
                {k^4\over \left( a + g' k^2 + \Gamma' k^4 \right) }
\label{o1}
\end{equation}

We evaluated the integrals appearing above numerically, and then fitted
experimental data\cite{bonart} on estane.
We had three free parameters to manipulate.
The three parameters at our disposal are $g$, a dimensionless diffusion
constant, $q$ the {\it pseudo charge} of our gauge theory, and finally
the length scale $\lambda$ which we used to turn the spatial lengths
in our theory dimensionless.
It is important to point out that the {\it form} of the 
renormalized structure function was crucial in obtaining reasonable
agreement with experiment.
The results of fitting Bonart's data are displayed in Fig. 4.
The values needed are listed in the figure caption.
It is worth noting that the value of the length scale $\lambda\sim 100 \AA$ 
we obtained
can be interpreted physically as the mean distance over which averaging has been
performed to go from an atomistic description to a mesoscale model. 
The volume described by this length scale can accommodate roughly $10^3$
polyurethane (hard segment) monomers.
With the current values of the parameters, our coupling constant $\alpha$
was just under 0.5.

Armed with values for our parameters appropriate for estane, 
we varied the concentration $c_h^0$ of the
hard segments of polyurethane in estane, decreasing it from Bonart's
value of 0.25.
In this way, we could probe how the location of the peak in the structure factor 
$\hat S(k)$ changed, as we varied the concentration of the hard segments.
The location of the peak is a measure of the inverse of the correlation
length in the system.
The purpose of this exercise was to investigate how the correlation length
behaves as the onset of self-assembly is approached.
Figure 5 shows a plot of the inverse correlation length as the concentration
is varied in the vicinity of $c^*$ 
(the minimum concentration below which self-assembly is impossible).
We found $c^*$ to be  approximately 0.1922 while
the critical exponent was found to be fairly close to 2/3 (0.6542).
We also found that as $c^*$ is approached,
the renormalized effective diffusion constant 
$g'_R = \left(g' + \delta g'(1) + \delta g'(2) + \delta g'(3)\right)  \to 0$,
implying that we have a precursor of critical slowing down.
Our calculation of this critical exponent
is the first which goes beyond the Gaussian approximation,\cite{riva} and
makes a prediction for diblock copolymers regarding the
the correlation length near the
onset of self-assembly, and shows that the onset
is accompanied by critical slowing down.
Experiments by Koberstein et al\cite{koberstein} on polyurethane-
polypropylene systems 
indicate qualitative agreement with our theory.
The difficulty in obtaining quantitative agreement lies in the
fact that experimental measurements of the structure factor
start to be swamped by the direct portion of the beam,
and the error bars increase dramatically,
as the peak marches towards the long wavelength limit.

%% file: sect4.tex
\section{ Conclusions}

We have shown that the concept of gauge invariance can be utilized
successfully at the mesoscale to generate the entropic/statistical
long-range forces responsible for self-assembly.
This was done by using local gauge invariance under the $SO(2)$ group
to derive the older (Gaussian) theories of self-assembly.
Our approach allows us to go naturally beyond the Gaussian approximation.
We computed the first nonvanishing contributions 
beyond the Gaussian approximation to the density-density correlation function
from the cubic and quartic terms in our energy functional.
We applied our theory to estane, a diblock copolymer, above its critical 
temperature.
We found that as the concentration approaches 
$c^*$ from above, the correlation length diverges
with a (2/3) power law, and the renormalized diffusion constant
tends to zero, implying critical slowing down.
The correlation function however, remains finite at the transition.
The divergence of the correlation length can be interpreted to mean that
as $c^*$ is approached from above,
the probability of finding mesoscale aggregates vanishes, so that
the average distance between aggregates diverges.
Experiments by Koberstein et al\cite{koberstein} on polyurethane-
polypropylene systems 
indicate qualitative agreement with our theory.
Universality arguments may be used to argue that our results
are applicable more generally.

We foresee a rich variety of applications of our approach to other
questions regarding diblock copolymers, such as 
their viscoelastic properties.
We also foresee investigations to time-dependent phenomena in copolymers,
such as detailed studies of critical slowing down at the onset of self-assembly,
and constitutive relations at non-zero strain rates.

%% file: sect5.tex
\section{ Acknowledgments}

I would like to acknowledge helpful comments from Carlo Carraro and
Michael Schick, and useful references provided by Frank Bates. 
Last but not least, I would like to thank Brian Kendrick,
Andrew Zardecki and Tony Redondo for useful discussions.
The work described in this paper was performed under the auspices of
the Los Alamos National Laboratory LDRD-CD project on polymer aging.

%% file: zfigs.tex
\begin{figure}
\caption{(a) is a pictorial representation of the cubic term in $\Delta a$.
Each leg corresponds to a factor of $c$, the field.
A $\partial_i$ indicates that a derivative of the field is to be taken
in the i-th direction.
A sum over i is understood.
The dark circle symbolizes a factor of $-\alpha \gamma$, 
the coupling constant.
The negative sign comes from the argument of the Boltzmann factor.
(b) is a pictorial representation of the quartic term in $\Delta a$.
A factor of $-\alpha$ is to be inserted at the intersection.}
\label{fig1}
\end{figure}
\begin{figure}
\caption{ (a) represents the {\it tadpole} diagram which is crucial in our
calculations.
(b) represents the {\it setting sun} diagram.
Both (a) and (b) are second order contributions to the correlation
function coming from the cubic interaction term, the first order
corrections being null.}
\label{fig2}
\end{figure}
\begin{figure}
\caption{ (a) and (b) represent 1-loop contributions from the quartic 
interaction term to the correlation term.}
\label{fig3}
\end{figure}
\begin{figure}
\caption{Comparison of experimental data and theory for estane.
The dark circles are experimental data points
without the background subtracted from them;
the solid curve is theory.
The experimental background is negligible, except for the data
point closest to the origin.
Both data and theory are normalized to the peak value.
The parameters used to obtain the fit were:
$c_h^0$=0.25, $g = 8000\AA^2$, $q = 0.6 {\times} 10^{-3}\AA^{-2}$,
and $\lambda = 95\AA$.
The peak in the experimental data and theory occurs at about
200 {\AA}, indicating the average distance between aggregates.
Note that the slight peak in the theoretical curve is obscured
by the experimental data.
It nevertheless exists and can be verified numerically.
The apparent disagreement between data and theory for large momenta is due
to the fact that the experimental background, which has not been
subtracted from the data, becomes important in this regime.}
\label{fig4}
\end{figure}

\begin{figure}
\caption{This is a plot of the location of the maximum in the theoretical
structure factor $S(k)$, as the concentration of the polyurethane
$c_h^0$ is varied in the neighborhood of the critical concentration c*,
which is $\approx 0.19216$. 
The open circles denote numerical results obtained in the manner
discussed in the text.
The solid line is the curve given by $0.71406 (c_h^0 - c^*)^{0.65424}$.
}
\label{fig5}
\end{figure}